\documentstyle[12pt]{article}
\def\np{\vfill\eject}       
\def\ns{\vskip 2pc\noindent}       
\def\cl{\centerline}

\def\part{\partial}
\def\siml{\underline{\sim}}
\def\ni{\noindent}
\def\IR{I\kern-.255em R}

\def\eps{\epsilon }

\def\kap{\kappa }
\def\th{\theta }

\def\Sibf{{\bf \Sigma}\, }

\def\Th{\hat{T}}
\def\Ttl{\tilde{T}}
\def\Ttlh{\hat{\tilde{T}}}
\def\kaptl{\tilde{\kappa}}
\def\pstl{\tilde{\psi}}
\def\thtl{\tilde{\theta}}
\def\thtlh{\hat{\tilde{\theta}}}
\def\kaptlh{\hat{\tilde{\kappa}}}
\def\kaph{\hat{\kappa}}

\def\Gmh{\hat{\Gamma}}

\def\thtlh{\hat{\tilde{\theta}}}

\def\psb{\overline{\psi}}

\def\thb{\overline{\theta}}

\def\Hbf{{\bf H}\,}

\def\Mbf{{\bf M}\,}

\def\Rbf{{\bf R}\,}

\def\yb{\overline{y} }

\def\lav{\vec{ \lambda}}

\def\np{\vfill\eject}       
\def\ns{\vskip 2pc\noindent}       
\def\ni{\noindent}
\def\IR{I\kern-.255em R}

\def\Gmbr{\overline{\Gamma} }
\def\kapbr{\overline{\kappa} }
\def\thbr{\overline{\theta} }
\def\Cbr{\overline{\bf C} }
\def\Tbr{\overline{T} }

\def\eps{\epsilon }

\def\Sibf{{\bf \Sigma}\, }

\def\Sbf{{\bf S}\, }

\def\Wbf{{\bf W}}

\def\eps{ \epsilon}

\def\Cbf{{\bf C}\,}

\def\Gbf{{\bf G}\,}

\def\Lbf{{\bf L}\,}
\def\Mbf{{\bf M}\,}
\def\Pbf{{\bf P}\,}
\def\Qbf{{\bf Q}\,}
\def\QB{{\bf Q_B}\,}
\def\Sbf{{\bf S}\,}

\def\dth{{dt \over 2}\,}
\def\Th{\hat{T}}

\def\la{\lambda}
\def\eps{\epsilon }

\def\gbf{{\bf {g}}}
\def\rb{{\bf {r}}}
\def\sb{{\bf {s}}}
\def\sbf{{\bf {s}}}
\def\ub{{\bf {u}}}
\def\ubf{{\bf {u}}}
\def\thb{{\bf {\theta}}}
\def\yb{{\bf {y}}}
\def\wb{{\bf {w}}}
\def\wbf{{\bf {w}}}
\def\ubh{\hat{\bf {u}}}
\def\ubb{\overline{\bf {u}}}
\def\ubt{\tilde{\bf {u}}}
\def\ubth{\hat{\tilde{\bf {u}}}}

\def\Bb{{\bf B}\,}

\def\Fb{{\bf F}\,}
\def\Fbf{{\bf F}\,}

\def\GB{{\bf \Bb}\,}

\def\Gbf{{\bf G}\,}
\def\Hb{{\bf H}\,}

\def\Jb{{\bf J}\,}
\def\Kb{{\bf K}\,}

\def\Pb{{\bf P}\,}

\def\Qb{{\bf Q}\,}
\def\Rb{{\bf R}\,}

\def\xbf{{\bf x}\,}
\def\Sib{{\bf \Sigma}\, }

\def\Sib{{\bf \Sigma}\, }

\begin{document}
\begin{center}
{\bf Optimal Estimation of Dynamically Evolving Diffusivities}
\end{center}
\begin{center}
{\bf Kurt S. Riedel \\
Courant Institute of Mathematical Sciences \\
New York University \\
New York, New York 10012}
\end{center}

\begin{abstract}
The augmented, iterated Kalman smoother is applied to system identification
for inverse problems in evolutionary differential equations.
In the augmented smoother, the unknown, time-dependent coefficients 
are included in the state vector, and have a
stochastic component. 
At each step in the iteration, the estimate of the time evolution
of the coefficients is linear.
We update  the  slowly varying mean temperature and conductivity by
averaging the estimates of the Kalman smoother.  
Applications include the estimation of anomalous diffusion coefficients
in turbulent fluids and the plasma rotation velocity  in plasma
tomography. 
\end{abstract}

\newpage
\noindent
{\bf 1. Introduction}

Estimation and system identification of distributed systems of 
partial differential equations (P.D.E's) are  
much researched fields. However, existing research in inverse problems
concentrates almost exclusively on the case where the system equations
are deterministic with unknown coefficients, and where
only the measurements have errors [3-5,22,26].
In contrast, finite dimensional estimation and control theory allows 
and stresses the importance of including stochastic forcing
in the system evolution equations to account for model error [15].  

Researchers attempt to model the effect of microscopic turbulence 
in plasmas and fluids with anomalous diffusion coefficients.
These effective equations for fluid flow are only an approximation
of the actual evolution equations, and in many cases the model error
is much larger than the measurement errors. Also, 
the anomalous diffusion coefficients are often time-dependent, while 
most research in inverse problems is restricted to time-independent
coefficients.
Model error has been included in the optimal estimation schemes of numerical
weather prediction [9,12]. 
By extending the state space, we are able not only to
estimate the state of the system, but also the coefficients. 

The Kalman filter-smoother is the optimal estimator of time-dependent state
vectors given noisy measurements and evolution equations with stochastic
forcing. The Kalman smoother minimizes a quadratic functional which
includes the residual squared error to the measurements. In addition, the 
Kalman filter generalizes the standard least squares analysis by including
a second term which is proportional to the square of the stochastic forcing.
The estimation problem is the mathematical dual to the control 
problem, and we refer the reader to Ref.~13 for an excellent description
of the computational aspects of fluid dynamical control. 

Since inverse problems are generally analyzed off-line, the 
fixed interval Kalman smoother is more appropriate than the Kalman filter.
In this article, we propose to analyze time-dependent inverse problems
with model error using a novel quasilinear 
extended Kalman smoother. To determine the unknown coefficients,
we extend the state space to include the original state space and
the coefficient space. 

As a running example, we consider the evolution of the  temperature
with an unknown, $time-dependent$ diffusivity,  $\kappa(\xbf,t)$:
$$\partial_t T  = \nabla\cdot\kappa(\xbf,t) \nabla T \
 + \sbf(t) \ +  \xi_1
\ ,\eqno (1.1a)$$
where $\sbf$ is the known source function. 
$\xi_{1}$ is a random field which stochastically forces the heat equation.
The random term represents small scale errors in the model.
We assume that the unknown diffusivity evolves as
$$\partial_t \thtl  = \mu_{2}\Delta \thtl + \xi_{2}
\ ,\eqno (1.1b)$$
where $\th= ln(\kappa(\xbf,t))$ and $\th\equiv \thbr(\xbf,t)+\thtl(\xbf,t)$
with known $\thbr(\xbf,t)$.

More generally, we consider stochastic evolutionary 
systems of the form: 
$$\partial_t \psi  = \Lbf(\th,t) \psi \ +\ \sbf \ + \GB \xi_{1}\ , $$
$$\partial_t \thtl  = \Mbf(t) \thtl + \xi_{2},
\eqno (1.2)$$
where $\psi$ is the state variable, and $\sbf$ is the known source function. 
The unknown parameter vector is $\th = \thb + \thtl$, where 
$\thbr(\xbf,t) \equiv {\rm E} [\th(\xbf,t)]$, and $\thtl \equiv \th - \thbr$.
$\Lbf(\th,t)$ and $\Mbf(t)$ are generators of smooth evolutionary 
semigroups. In our case, $\psi$ is the temperature field,
$\psi = T(\xbf,t)$. 
$\xi_{1}$ and $\xi_{2}$ are random fields 
which represent the stochastic forcing. The dimension of the noise, $\xi$,
can be smaller than the dimension of $(T,\thtl)^*$.
$\GB$ is a bounded linear operator which propagates the noise.


Equations (1.1) and (1.2) are quasilinear stochastic differential equations
(s.d.e.'s)
 with a particular upper triangular structure. 
We solve the nonlinear s.d.e's  iteratively, by linearizing
the random component of Eqs. 1.1-2 about our current estimate 
of its expected value and using a quasilinear closure for the mean field.  
We define $\psb(\xbf,t) \equiv {\rm E} [\psi(\xbf,t)]$, 
and $\pstl \equiv \psi - \psb$ . The linearized evolution equations
for the fluctuations are
$$\partial_t \pstl  = \Lbf(\thbr,t) \pstl + \Gbf(\thbr,t)\thtl + 
\GB \xi_{1} \ ,\eqno (1.3a)$$
$$\partial_t \thtl  = \Mbf(t) \thtl + \xi_{2},
\eqno (1.3b)$$
where 
$\Gbf(\thbr,t)\equiv {\partial \Lbf(\th,t)\psb \over \partial \th}(\thbr)$.
We neglect the nonlinear terms in the fluctuating amplitudes
for the fluctuation equation.
For the inverse heat conductivity problem of Eq.~(1.2), Eq.~(1.3) reduces to
$$\partial_t \Ttl  = \nabla\cdot \kapbr(\xbf,t) 
[ \nabla \Ttl +  \thtl \nabla \Tbr ] +  \xi_1 \ ,\eqno (1.4a)$$
$$\partial_t \thtl  = \mu_{2}\Delta \thtl + \xi_{2} 
\ .\eqno (1.4b)$$
Equations (1.3-4) describe the stochastic component of $\psi$ and $\th$.
When $\Tbr$ and $\kapbr$ are given, {Eq's~(1.4)} are a standard 
distributed estimation problem as discussed  in [6,25].
When $\Tbr$ and $\kapbr$ are unknown, one approach is to minimize
the residual sum of squares as a function of $\Tbr$ and $\kapbr$
as well as $\Ttl$ and $\thtl$ with a smoothness  penalty function [1,6].
This minimization is numerically difficult. In Sec.~V,
we describe a different approach where we modify the standard iterated
Kalman filter by 
updating $\Tbr$ and $\thbr$ with the slowly varying part of the estimates of
$\Ttl$ and $\thtl$.

In Section II 
and  Appendices A \& B, we review the Kalman filter-smoother.
In Appendix A, we  give variational formulations which may be 
useful for more advanced time discretizations. Appendix C describes the
competing penalized least squares approach and gives a simple hybrid model.

In Section III, we discuss discrete approximations to the distributed
estimation problem and regularizations of them. 
We describe  a numerical implementation of the Kalman smoother 
for a one-dimensional heat equation with an unknown time-dependent
diffusion coefficient.
{ In our  
augmented  Kalman smoother formulation, a number of quantities need to
be specified {\it a priori}: the covariances of $\xi_1$ and $\xi_{2}$,
the operator, $\Mbf(t)$, and the strength of the smoothness penalty.}
In Section IV,  we examine the selection  of these {\it a priori} terms.  
Appendix E relates 
the $a$ $priori$ smoothness of $\Ttl$ and $\kaptl$ to $\Mbf$ and 
the covariances of $\xi$ and $\xi_{2}$.  
In Section V, we update the Kalman smoother by adding
the slowly varying part of the filter estimates  
to $\Tbr$ and $\thbr$.
In Section VI, we discuss applications to plasma physics  
and to fluid dynamics. 


\np
{\bf II. Infinite Dimensional Filtering of Distributed Systems}

\medskip

We briefly summarize the similarities and differences for estimating
infinite dimensional systems. For more rigorous treatments, 
we recommend [6,10,25]. 
Appendix B contains a more explicit review of the finite 
dimensional, discrete time case. 
We rewrite Eq's~(1.4) using the augmented state vector,
$\ubf^{*} = (T,\thtl)^{*}$:
$$
\partial_t{\ub}(t) = {\bf F} (t) \ub(t) + {\bf \GB} (t) \xi_u (t)
,\eqno (2.1)$$
where $\xi_u^{*} = (\xi_{1},\xi_{2})^{*}$, 
with $E[ \xi_u (t) \xi_u^{*} (s)] = \Qb(t) \delta (t-s)$
and 
$$ \Fb_t \equiv \left( \begin{array}{c}
 \nabla \cdot \kapbr(\xbf,t)  \nabla \\ 0 \end{array}
\begin{array}{c} 
{ \nabla \cdot   \Gmbr(\xbf,t) } \\  \mu_2 \Delta 
\end{array} \right) ,\eqno (2.2)$$
where $\Gmbr(\xbf,t)\equiv \kapbr(\xbf,t) \ \nabla \Tbr$ is the mean
heat flux.
To properly treat the infinite dimensional case,
we assume that $\Qbf_u(t)$ is  a positive definite trace class covariance,
and that $\Fb(\thbr(t),t)$ generates a smooth evolutionary semigroup.
The noise propagator, $\GB$, is useful because $\GB \Qbf_u \GB^*$
can be semi-definite.


In this section, we assume that the measurements are
continuous in time and satisfy
$$
{\yb}(t) = \Hb (t) \ub(t) + \eps (t)
,\eqno (2.3a)$$
where  $\eps(t)$ is the measurement error
and its covariance is $E[ \eps (t) \eps^{*} (s)] = \Rb(t) \delta (t-s)$.
$\Hb$ is a {\it bounded} linear map. In our example, $\yb(t)$ consists of the 
measured temperature values at $m$ distinct locations, $r_{\ell}$, with
sampling times $t_i$,
where $i =1,\ldots t_f$ and $\ell =1, \ldots m$. 
The measurements are assumed to be spatially and temporally independent:
$$
T_{\ell, i} = T(\xbf_{\ell} ,t_i ) + \epsilon_{\ell i} \ , \ \
{\rm cov} ( \eps_{\ell, i},\eps_{k,j} ) = \sigma_k^2 
\delta_{\ell, k} \delta_{i,j} \ .
\eqno (2.3b)$$
In a basis function representation, $\ubf(\xbf,t)=\sum_{k} u_k(t)g_k(\xbf)$,
and $\Hbf$ has the representation: $H_{\ell,k}=g_k(r_{\ell})$. 
Equation (2.3a) allows for chordal cross-section measurements 
since ${y}_{\ell}(t) = H_{\ell} (t) \ub(t)$
can be rewritten as ${y}_{\ell}(t) = \int H_{\ell}(\xbf,t) \ub(\xbf,t)\ d\xbf$.

If the measurement times are fast relative to the characteristic
evolution time, the optimal filter will be well modeled by the continuous
measurement time model. 
The continuous measurement time filter is the limit of 
the discrete measurement time filter with a measurement covariance of
$\Rbf_t = {\Rbf(t) \over \Delta}$, where $\Delta$ is the time
between measurements. In this limit, the Kalman filter simplifies
because the  measurement covariance dominates the filter covariance.

We suppress the time-dependence of $\Hb(t)$, $\GB(t)$,  
$\Qb(t)$, and  $\Rb(t)$ in this section for notational
simplicity.  
The Kalman filter satisfies
$$ \partial_t
\hat{\ub}(t) = \Fb_t \hat{\ub} (t) + \Pb(t) \Hb^{*}
\Rb^{-1} ({\yb} - \Hb \ub ) \ , \eqno (2.4)
$$
where $\hat{\ub}$ denotes the estimate of  $\ub$. The covariance of
$\hat{\ub}$ evolves as
$$
\partial_t{\Pb}(t)=\Fb_t\Pb(t) + \Pb(t) \Fb_t^{*} + \QB - 
\Pb\Hb^{*} \Rb^{-1} \Hb\Pb 
\ , \eqno (2.5)$$
where $\QB \equiv \GB \Qb \GB^{*}$. 
The initial conditions are 
${\rm E}[\ubf]{\ } (t=0) = {\ubf}_0$ and $\Pb(t=0) = {\Pb}_0$. 
$\Pb^{-1}(t)$ evolves as well: 
$$
\partial_t{\Pb}^{-1}(t)= - \Pb^{-1}\Fb_t - \Fb_t^{*}\Pb^{-1}  
- \Pb^{-1}  \QB \Pb^{-1}  + 
\Hb^{*} \Rb^{-1} \Hb \ . \eqno (2.6)
$$
For the continuous time fixed interval smoother, we integrate backwards
in time the stabilized equations:
$$
\partial_t{\ubh}(t|t_f) = \Fb_t\ubh(t|t_f)  
+ \QB\Pb^{-1}(t) 
\left[ \ubh(t|t_f)- \ubh(t|t) \right]
,\eqno (2.7)$$
{\cl{$\partial_t{\Pb}(t|t_f) = 
[\Fb_t + \QB\Pb^{-1}(t)] {\Pb}(t|t_f)   + 
{\Pb}(t|t_f)  [\Fb_t + \QB\Pb^{-1}(t)]^{*} 
$}}
$$ - \QB -
{\Pb}(t|t_f) \Hb^{*} \Rb^{-1} \Hb {\Pb}(t|t_f) 
,\eqno (2.8)$$
with the final conditions $\ubh(t_f|t_f) =\ubh(t_f) $, 
$\Pb(t_f|t_f) = \Pb(t_f)$. The backwards time integration is normally 
ill-conditioned. To remedy this,
the Bryson-Frazier (B.-F.) formulation [8] of the continuous time smoother 
uses the auxiliary variables, $\gbf (t)$ and $\Gbf(t)$, where
$$
\hat{\ub} (t|t_f ) = \hat{\ub} (t) - \Pb(t) \gbf (t) \ ,\eqno (2.9)
$$
$$
\Pb(t|t_f ) = \Pb(t) + \Pb(t) \Gbf (t) \Pb(t) \ . \eqno (2.10)
$$
The auxiliary variables satisfy
$$
\partial_t{\gbf} (t) = -[\Fb_t - \Pb(t) \Hb^{*} \Rb^{-1} \Hb ]^{*} \gbf +
\Hb^{*} \Rb^{-1} [ {\yb}_t - \Hb \hat{\ub} (t) ] \ ,  \eqno (2.11)
$$
$$\partial_t
{\Gbf} (t) = -[\Fb_t - \Pb(t) \Hb^{*} \Rb^{-1} \Hb ]^{*} \Gbf -
\Gbf [\Fb_t - \Pb(t)\Hb^{*} \Rb^{-1} \Hb] + 
\Hb^{*} \Rb^{-1} \Hb \ ,\eqno (2.12)$$
with the final conditions $\gbf (t_f) = 0$, $\Gbf(t_f) = 0$.
The B.-F. formulation replaces $- \Fb_t$ with 
$\Fb_t - \Pb(t) \Hb^{*} \Rb^{-1} \Hb$, and thereby tends to stabilize
the backward time integration. Although the B.-F. formulation is common
in finite dimensional control, {\it we are unaware of any previous usage
in distributed systems.} 

\ns \noindent
{III. \bf Discrete Approximations of Distributed Systems} 

\medskip

To apply the Kalman smoother to distributed systems of partial differential
equations, we represent the system using a truncated set of basis functions,
and discretize the time evolution of the augmented system with a stable, 
consistent time advance. We assume that the augmented evolution equations
are well-posed, and constitute a strongly continuous semigroup.
The Lax equivalence theorem [24] 
implies that any stable, consistent discretization converges 
to the continuous time
limit. Similarly, the covariance evolution is discretized with a stable,
consistent numerical scheme. We replace Eqs. (2.7-8)  with the stabler 
Bryson-Frazier formulation, (2.9-12). 

We now examine the inverse heat diffusivity problem in the one-dimensional
periodic case. As discussed in Appendix C, the inverse problem is 
ill-conditioned, and we regularize the Kalman smoother 
by adding a small higher-order spatial dissipation operator (hyperdiffusion).
The hyperdiffusion damps the small scale oscillations 
and thereby aids in convergence. 
Furthermore, we remain within the standard framework of Kalman filters.
Section IV addresses the selection of the size of the hyperdiffusivity 
coefficient. Thus we replace Eq.~(1.4a) with
$$
{\partial_t \Ttl  } \ (\xbf,t) = 
{\partial_x } \left[\kapbr(\xbf,t) \ {\partial_x \Ttl } \ +
 \Gmbr(\xbf,t) \ { \thtl } \right]
\ - \mu_1 \part_x^4 \Ttl + S(\xbf,t)+ w(\xbf,t)
\ , \eqno (3.1)$$
where $\Gmbr(\xbf,t)\equiv \kapbr(\xbf,t) \ \partial_x \Tbr$ is the mean
heat flux.
$S(\xbf,t)$ is a known source term, and $w(\xbf,t)$ is random forcing.
We assume periodic boundary conditions: $T(\pi,t) =  T(-\pi,t)$ 
and $\partial_x T(\pi,t) = \partial_x T(-\pi,t).$ 
We expand both the mean quantities,
$\Tbr(\xbf,t)$ and $\Gmbr(\xbf,t)$ and the fluctuating quatities
$\Ttl(\xbf,t)$ and $\thtl(\xbf,t)$ in truncated Fourier series:
$$T(\xbf,t) = \sum_{k=-N_T}^{N_T} T_k(t)e^{ikx} \ , \ 
\Gmbr(\xbf,t) = \sum_{k=-N_T}^{N_T} \Gmbr_k(t)e^{ikx} \ , \ 
\thtl(\xbf,t) = \sum_{k=-N_T}^{N_T} \thtl_k(t)e^{ikx}
\ ,\eqno (3.2)$$
where $T_k(t)$, $\Gmbr_k(t)$ and  $\th_k(t)$ are complex with 
$T_{-k}(t)=T_k(t)^*$, $\Gmbr_{-k}(t)=\Gmbr_k(t)^*$ and  
$\thtl_{-k}(t)=\thtl_k^*(t).$ 
The nonlinear transformation between $\kap(\xbf,t)$ and $\th(\xbf,t)$
is performed by collocation at the spatial points, $x_k= {\pi k \over N_t}, \ $
$k =-N_T \ldots N_T$ using the Fourier transform.

In Fourier space, the diffusion equation becomes
$$
{d \Ttl_k \over dt } \ (t) = \ 
-\sum_{k'=-N_T}^{N_T} kk' \kapbr_{k-k'} \Ttl_{k'} 
-\sum_{k'=-N_T}^{N_T} k \Gmbr_{k-k'} \thtl_{k'} 
-\mu_1 k^4 \Ttl_k(t) + S_k(t)+ \xi_k(t) \ .\eqno (3.3)$$
We assume that the stochastic forcing of the different modes is statistically
independent, and decays algebraically:
${\rm Exp}[\xi_{k}(t){\xi}_{k'}^*(t')]= 
\alpha_1 |k|^{-\beta_1} \delta_{k,k'}\delta(t-t').$ 
Our model for the stochastic evolution of $\th_k$ is
$$
{d \thtl_k \over dt } \ (t) = \ - \mu_{2} |k|^{2} \thtl_{k}(t)
+ \xi_{2,k}(t) \ , \eqno (3.4)$$
where ${\rm Exp}[\xi_{2,k}(t){\xi}_{2,k'}^*(t')]= \alpha_{2}|k|^{-\beta'}
\delta_{k,k'}\delta(t-t').$
In Sec.~IV, we discuss the selection of
the free parameters such as $\alpha_1,\ \alpha_{2},\ \beta$ and $\beta'$ .
For the time discretization, 
the stochastic forcing is scaled as the square root of the time step size:
$\xi_k \equiv {d w_k\over dt}$ with 
$${\rm Exp}[w_{k}(t){w}_{k'}^*(t')]= \alpha_{1}|k|^{-\beta_1}
\delta_{k,k'}  dt \ , \ 
{\rm Exp}[w_{2,k}(t){w}_{2,k'}^*(t')]= \alpha_{2}|k|^{-\beta_2}
\delta_{k,k'}   dt \ 
\eqno (3.5).$$
Thus the Brownian increment, $w_k$, is large relative to the time step.
As a result, the numerical accuracy of the finite difference approximation
of the  s.d.e. can be much worse than the accuacy of the same scheme on 
deterministic differential equations [21]. Recently, higher-order difference
schemes have been developed, and we use the Milstein implicit 
second-order weak Taylor scheme (pg. 499 of Ref.~21). 
This scheme is globally second-order accurate
for computing weak solutions, but it is only first-order accurate for 
computing strong/pathwise solutions. Since we are primarily interested in 
estimating the mean quantities, $\Tbr(\xbf,t)$ and $\kapbr(\xbf,t)$,
 weak convergence is adequate. The implicit Milstein discretization of
Eqs. (3.3-4) is  
$$
{\thtl_k(t+dt) } \  =
{ 1 -\dth \mu_{2} |k|^{2}\over 
1 +\dth \mu_{2} |k|^{2} } \thtl_{k}(t)
\ + \ {w_{2,k} (t)\over 1 +\dth \mu_{2} |k|^{2} } 
\ ,\eqno (3.6)$$

$$ {\Ttl_k(t+ dt) } \ = \Ttl_k(t) \ + \
dt \sum_{k'=-N_T}^{N_T} kk' \kapbr_{k-k'} 
\left({\Ttl_{k'}(t)+ \Ttl_{k'}(t+dt) \over 2}\right)
$$ $$ 
+ dt \sum_{k'=-N_T}^{N_T} k \Gmbr_{k-k'} 
\left({\thtl_{k'}(t)+ \thtl_{k'}(t+dt) \over 2}\right)
+ S_k(t+\dth)dt + w_k(t)  \ .\eqno (3.7)$$
The term, 
$\sum_{k'=-N_T}^{N_T} kk' \kapbr_{k-k'} \Ttl_{k'}(t+dt) $,
makes Eq.~(3.7) intractable, and therefore we expand Eq.~(3.7) assuming
the average diffusion, $\kapbr_o$, is large relative to the spatial
variation of $\kapbr$. 
Using this semi-implicit approximation [23], the temperature time advance 
becomes 
$$
{\Ttl_k(t+ dt) } \ = \ 
f_k \Ttl_{k}(t) -
\sum_{\stackrel{k'=-N_T}{k'\ne k}}^{N_T} g_{k,k'}  \Ttl_{k'} \ +
\sum_{k'=-N_T}^{N_T} h_{k,k'} \thtl_{k'} \ +
$$ $$
\sum_{k'=-N_T}^{N_T} c_{k,k'} w_{2,k'}
+ S_k(t+\dth)dt + w_k(t)  \ ,\eqno (3.8)$$
where $d_k \equiv 1 + \dth (\kap_0 |k|^{2} + \mu_1 |k|^4 )$ and
$$f_k \equiv {1 - \dth (\kap_0 |k|^{2} + \mu_1 |k|^4 )  \over  
1 + \dth (\kap_0 |k|^{2} + \mu_1 |k|^4 )} \ , \ 
\ g_{k,k'}\equiv {kk' \kapbr_{k-k'}dt \over 2} 
\left( 1 +  {1 \over d_k d_{k'}} \right)
$$

$$
h_{k,k'}\equiv {k \Gmbr_{k-k'}dt \over 2 d_k} 
\left(1 + { 1 -\dth \mu_{2} |k'|^{2}\over 
1 +\dth \mu_{2} |k'|^{2} }
\right)\ \ , \
 c_{k,k'}\equiv {k \Gmbr_{k-k'}dt \over 2 d_k 
(1 +\dth \mu_{2} |k'|^{2} )}
\ .$$

Note that $w_{2,k}$ appears in Eq.~(3.8) due to the difference scheme and
this term is included in the Kalman smoother via the propagator matrix,
$\Bb$, in Sec.~II and Appendix B.  
In Eq.~(3.8), only the nonlinear evolution of the nonlinear terms 
is first-order in time while the other terms are second-order accurate. 
The partially implicit difference
scheme adds extra dissipation.

The observational data is transformed to mode space:
$y_k(t_i) = \sum_{\ell=1}^{m} y_{\ell,i}e^{-ikx_{\ell}}$
where $|k| \le {m \over 2}$. $y_k(t_i)$ is complex with 
$y_{-k}(t_i)= {y}_k^*(t_i).$ When $m$ is even, we omit the $cosine$ component
of the last coefficient, $k= {m\over 2}$. 
In mode space, the measurement error matrix is 
$R_{k,k'} = \sum_{\ell=1}^{m} \sigma_{\ell}^2 exp(i(k'-k) x_{\ell})$
where $0\le k,k' \le {m \over 2}.$
The measurement evaluation matrix is 
$H_{k,k'} = \sum_{\ell=1}^{m}  exp(i(k-k') x_{\ell})$ where 
$0\le k \le {m \over 2}$ and $|k'| \le N_T$.
When $\sigma^2_1=\sigma^2_2= \ldots =\sigma^2$ and the measurement
locations are uniformly distributed, $x_{\ell}= \pi{(2\ell-m-1)\over m}$, 
the measurement
error matrix reduces to $R_{k,k'} = m \sigma^2 \delta_{k-k'}$,
and the evaluation matrix reduces to $H_{k,k'} = m $
when ${k-k'}=0$ mod $m$ and zero otherwise.

Equations (3.6) \& (3.8) constitute a discrete dynamical system.
We assume that the measurement times occur on the time scale
of a single time step advance of the diffusion equation,
and  use the discrete measurement time filter-smoother which
is described in Appendix B. The time discretized equations  have now 
been placed in the block tridiagonal form of Eq.~(B11). 
Our preferred numerical method to solve Eq.~(B11) is sequential quadratic
programming [11,29]. Alternatively, we can solve Eq.~(B12)
using conjugate gradient iterations on the diagonal subblocks. 


 \ns \noindent
{\bf IV. Models for Stochastic Parameter Evolution}

\medskip

Ideally, the model  error  covariances, $\Qb_{\GB}$ and $\Qb_{\th}$,
are given $a$ $priori$ or are estimated from the residuals.
We denote by $\Cbf(t,s)$ the covariance of 
$\ubf^{*} = (\Ttl,\thtl)^{*}$ in the {\it absence of measurements}.
($\Pbf(t,s)$ is the covariance of the estimates, $\ubh$ {\it conditional
on the measurements.}) In practice, 
we often have  better knowledge of 
$\Cbf_{}$ than $\Qbf_{}$.
Appendix E expresses $\Cbf_{}$ in terms of $\Qbf_{}$ for time-independent
evolution equations.

To explicitly evaluate an approximate covariance, we neglect the spatial
temporal variation in $\kapbr(\xbf,t)$ 
and make a large $k$ expansion of Eq.~(E4).
In this case, the evolution of $T_k$ and $\th_k$ decouple to leading order
and the quasistationary approximate covariance satisfies
$$C_{\th,k,k}(t) \ \siml \ {Q_{\th,k}(t) \over 2\mu_{2} |k|^{2} }\ \ \ , \ \ \ 
C_{T,k,k}(t)\  \siml \ {Q_{T,k,k}(t) \over 
2(\kapbr |k|^{2} + \mu_{1} |k|^{4})} \ 
\ . \eqno (4.1)$$
These balance equations for the  evolution
of  $C_{T,k}$ and  $C_{\th,k}$ can be used to define values of
$\alpha_{T}$, $\alpha_{\th}$, $\beta$ and $\beta'$, provided that
$C_{T,k}$ and  $C_{\th,k}$ are given.

To choose $\mu_1$, we neglect $\part_t T_k$ in Eq.~(3.3) and assume that
$\kappa(\xbf,t)$ is spatially independent with value $\kappa_0$
. In this case, we have
$\Th_k(t|t=0) = S_k(t)/(\kappa_0 k^2 +\mu_1 k^4)$, which has a bias error
of $-{\mu_1\over \kappa_0} k^2S_k(t)/(\kappa_0 k^2 +\mu_1 k^4)$ 
and a variance of
$Q_{T,k}(t)/(\kappa_0 k^2 +\mu_1 k^4)^2.$ Thus the total expected  error for
is approximately equal to
$$\sum_k{ Q_{T,k}(t)+|{\mu_1\over \kappa_0} k^2S_k(t)|^2
\over ( \kappa_0 k^2 + \mu_1 k^4 )^2}\  . \eqno(4.2) $$
The size of the regularizing term, $\mu_1 \Delta \Delta$, 
is chosen to minimize this expected error. 

The initial conditions are unknown, and so we include the initial 
conditions in the iteration. Our initial guess is $\kap(\xbf,t=0) = \kap_0$
and $T_k(t=0) = S_k/\kap_o k^2$ with $T_0(t=0)$ chosen by dimensional
considerations:  $T_0 = S_{rms}/\kap_o $ where $S_{rms}$ is the root mean
square average of $S(\xbf,t)$.
We also choose a smoothness prior for $\Pb(t=0|0)$: $\Pb(t=0|0) = 
(S^o_k/\kap_o k^2)^2$ where $S^o_k \equiv {\rm max}\{ S_k(t=0) , S_{rms}\}$.

\ns \noindent
{\bf V. Temporal Averaging to Update $\Tbr(\xbf,t)$ and $\thbr(\xbf,t)$}

\medskip

In Eqs. (1.3-4), we separated the temperature into a mean field, 
$\Tbr(\xbf,t)$,
and a fluctuating field, $\Ttl(\xbf,t)$, which we estimate with the Kalman
smoother. In practice, $\Tbr(\xbf,t)$ and $\thbr(\xbf,t)$ are unknown, and we
use an iterative procedure to estimate them. We let 
$\Tbr^{(\ell)}(\xbf,t)$ and $\thbr^{(\ell)}(\xbf,t)$ 
denote the $\ell$th iterate of the mean field and $\Ttlh^{(\ell)}(\xbf,t)$ and 
$\thtlh^{(\ell)}(\xbf,t)$ 
denote the Kalman smoother estimate of $\Ttl(\xbf,t)$ and $\thtl_{}(\xbf,t)$ 
when the filter is linearized about 
$\Tbr^{(\ell)}(\xbf,t)$ and $\thbr^{(\ell)}(\xbf,t)$.
The model misspecification on the $\ell$th iterate,
$\Tbr^{(\ell)}(\xbf,t)-\Tbr_{}(\xbf,t)$ and $\thbr^{(\ell)}(\xbf,t) - \thbr_{}(\xbf,t)$,  
is modeled by the Kalman smoother as part of the stochastic forcing 
in the model. 
Thus the Kalman smoother implicitly includes  and corrects for the
possibility of model misspecification in its estimator.
A simple iteration scheme would be 
$$\Tbr^{(\ell+1)}(\xbf,t) = \Tbr^{(\ell)}(\xbf,t) \ + \
{\rm E}[ \Ttlh^{(\ell)}(\xbf,t)]
\ ,\eqno(5.1a)$$ 
$$\thbr^{(\ell +1)}(\xbf,t) = \thbr^{(\ell)}(\xbf,t) \ + \
{\rm E}[ \thtlh^{(\ell)}(\xbf,t)]
 \ . \eqno(5.1b)$$ 
{\it However, $\Tbr^{(\ell+1)}(\xbf,t)$, estimated from Eq.~(5.1a), 
is not the solution of the heat equation with the diffusion
coefficient, $\kapbr^{(\ell+1)}\equiv \exp[\thbr^{(\ell+1)}(\xbf,t)]$, 
from Eq.~(5.1b).} Therefore, we replace the naive $\Tbr$ update
with a heat flux averaging update. 

In our heat flux averaging implementation
of the mean temperature update, we define the 
$(\ell +1)$th estimate of the total heat flux 
by $\Gmh^{(\ell+1)} \equiv \left[ \kapbr^{(\ell)}+ \kaptlh^{(\ell)}\right]
\nabla \left[ \Tbr^{(\ell)}+ \Ttlh^{(\ell)}\right]$.
The mean heat flux is the expectation of the nonlinear heat flux:
$$\Gmbr^{(\ell)}(\xbf,t) = 
{\rm E}[ \Gmh^{(\ell)}(\xbf,t)] \ . \eqno(5.2)$$ 
We evolve the mean temperature with the mean heat flux:
$$
{\partial_t \Tbr^{(\ell)} } \ (\xbf,t) = 
\nabla \cdot \Gmbr^{(\ell)}(\xbf,t) 
\ - \mu_1 \Delta\Delta \Tbr^{(\ell)} + S(\xbf,t)
\ . \eqno (5.3)$$
Given $\Gmbr^{(\ell)}$ and $\Tbr^{(\ell)}$ from Eqs. (5.2-3), we solve for
$\kapbr^{(\ell)}(\xbf,t)$ using
$$\nabla\cdot \kapbr^{(\ell)}\nabla \Tbr^{(\ell)}(\xbf,t) =
\nabla \Tbr^{(\ell)}\cdot \nabla \kapbr^{(\ell)}(\xbf,t)
\ + \ \kapbr^{(\ell)}(\xbf,t)\Delta \Tbr^{(\ell)}
=\Gmbr^{(\ell)} (\xbf,t) \ .\eqno (5.4)$$
Since $\Tbr^{(\ell)}(\xbf,t)$ is given, Eq.~(5.4) is a linear hyperbolic
equation for $\kapbr^{(\ell)}(\xbf,t)$. Equation (5.4) is well-posed,
provided that
$\Tbr^{(\ell)}$  does not vanish in the  domain and that all the 
$\nabla\Tbr^{(\ell)}$ characteristics intersect the boundary,
where boundary data  for $\kapbr$ are given.

We implement  the expectation operator by  averaging
$\nabla \cdot \Gmh^{(\ell+1)}(\xbf+\xbf',t)$ with a smoothing
kernel:
$$\nabla\cdot \Gmbr^{(\ell+1)}(\xbf,t) = 
\int K(\xbf',t')\nabla \cdot \Gmh^{(\ell+1)}(\xbf+\xbf',t')dt'd\xbf'
\ ,\eqno(5.5)$$  
where $K(\xbf,t)$ is a smoothing kernel 
with a characteristic
duration of three to five autocorrelation times. 
In the periodic heat equation of  Sec.~III, we smooth the $k$th harmonic
of the heat flux, $\Gmh_k$,
with a kernel of duration $1/ \kapbr |k|^2$.
The longer the kernel window, the more the high frequency oscillations
are suppressed. 
We average the divergence of the heat flux instead of the heat flux
because the  divergence of the heat flux enters into Eq.~(5.3).


In Sec.~III-V, we have described a detailed model  for using the Kalman 
smoother to estimate  $\kappa(\xbf,t)$ including stochastic dynamics.
We couple the numerical discretization  of Sec.~III with the covariance
model of  Sec.~IV.  Sec.~V  completes the iteration  cycle by updating
$\Tbr(\xbf,t)$ and $\kapbr(\xbf,t)$. Our heat flux averaging update 
eliminates the nonlinear discrepency, ${\rm E} [ \nabla\cdot  \kaptlh 
\nabla \Ttlh] - \nabla\cdot  {\rm E} [\kaptlh]  \nabla {\rm E}[\Ttlh]$,
from the updating scheme.
We now sketch several more realistic 
inverse problems for the augmented Kalman smoother.

\ns \noindent
{\bf VI. Potential Applications}

\medskip

A) Elliptic Identification Problems

Coefficients for elliptic equations may be identified using the
augmented Kalman smoother by adding a vanishingly small time derivative to the
equations [17]. 
Similarly, when the temperature is time  dependent and coefficients
are time-independent, we can regularize the diffusivity evolution  equation
with a vanishingly small time derivative of the parameters, $\eps \part_t \th$.

\medskip 

B) Anomalous Transport Coefficients in Fusion Plasmas
\medskip

In magnetic fusion, high temperature plasmas are confined in
diffusive equilibrium by external magnetic fields. Due to the presence
of low-level turbulence, the transport of heat and particles is
anomalously large.
We consider the important problem of estimating  anomalous
transport coefficients in fusion plasmas. 

To good approximation, fusion
plasmas are determined by the electron and ion temperatures, $T_e
(r,t)$, $T_i (r,t)$, and the electron and ion densities, $n_e (r,t)$ and
$n_i (r,t)$. 
A simple, but realistic set of power balance equations are
$$
{3 \over 2} \ {\partial( n_e T_e )\over \partial t} \ (r,t) =
{3 \over 2}\nabla \cdot (\chi_e n_e \nabla T- n_e TV) - n_e T_e\nabla \cdot V
+ S_e (r,T,n) + n_e \
{T_i - T_e \over \tau_{ei} }
\ , \eqno (6.1)$$
$$
{\partial n_e \over \partial t} = \nabla \cdot (n_e V + D_e \nabla n_e )
\ , \eqno (6.2)$$
$$
{3 \over 2} \ {\partial (n_i T_i) \over \partial t} \ (r,t) =
{3\over 2}\nabla \cdot (\chi_i n_i \nabla T_i - n_e TV) - n_i T_i\nabla \cdot V
+ S_i (r,T,n) - n_e \
{T_i - T_e \over \tau_{ei} }
\ , \eqno (6.3)$$
$$
n_e = Z_{}n_i
\ . \eqno (6.4)$$
Equations (6.1-4) consist of three parabolic equations for $n_e$, $T_e$ and
$T_i$ with a constraint to determine $n_i$. There are four unknown
transport coefficients: the electron and ion diffusivities, $\chi_e
(r,t)$, $\chi_i (r,t)$, the electron density diffusivity, $D(r,t)$, 
the pinch velocity, $V (r,t)$.
The source terms, $S_e (r,T,n)$ and $S_i (r,T,n),$ and the
coupling coefficient, $\tau_{ei},$ depend only weakly on the unknown profiles.
For convenience, we neglect these nonlinearities in the Kalman smoother.
$Z$ is the  change number. 
Similar sets of transport equations have been used for thermal control
of fusion plasmas [7,14,18,20,26,28]. 

Presently, two algorithms are applied to estimate the transport coefficients.
When the diffusion coefficients are specified using a low-order parametric
models, least squares can be used to estimate the unknown free parameters
[26,28]. 
A second nonparametric approach is to estimate
the measured profiles, $n_e(r,t)$, $T_e(r,t)$, and $T_i(r,t),$ 
by smoothing the raw data in space and time, and then inverting Eq.~(6.1-4)
for the transport coefficients [14]. 
The smoothing of the raw measurements will bias the estimated transport 
coefficients to higher values.

The augmented Kalman smoother is ideally suited to estimate the
state variables, $n_e(r,t)$, $T_e(r,t)$, and $T_i(r,t)$, and
the time-dependent transport coefficients, given point measurements
of the state variables. Furthermore, the augmented Kalman smoother
produces realistic covariance estimates for the diffusion coefficient
including the sizable model error in Eqs. (6.1-4).


\medskip

C) Turbulent closures of fluid equations
\medskip

Another application is to 
estimate the effective equations for low level fluid turbulence. 
The simplest variant of this problem is to assume the fluid velocity,
$\ub(\rb,t)$, satisfies the Navier-Stokes 
equations: 
$$\part_t \ub + \ub\cdot \nabla \ub =\nabla p + \nu\Delta \ub +\xi \ ,$$ 
$$\nabla \cdot \ub= 0 \ .$$
The stochastic forcing, $\xi(\rb,t)$, models the fine scale fluid turbulence. 
The viscosity, $\nu(\rb,t)$, is anomalously large to account for the
macroscopic response of the fluid to the subgrid scale turbulence. 
To model the  evolution of $\nu(\rb,t)$, we assume that
$\th(\rb,t)=ln(\nu(\rb,t))$ is convected and stochastically forced:
$\part_t \thb + \ub\cdot \nabla \thb =\xi.$

We are given continuous time measurements of velocity field 
on a coarse grid in space. We expand the Navier-Stokes equation in
the set of eigenfunctions of the laminar flow linear stability problem,
and apply the augmented Kalman smoother to the problem. 
In imposing a higher-order spatial dissipation for additional numerical
stability, we note that many turbulence theories actually contain a
hyperviscosity.


\np

D) Time-dependent tomography of plasma instabilities
\medskip

X-ray tomography is used to analyze the $m=1$ instability in tokamak 
plasmas [16]. Presently, each time slice is inverted separately to construct 
a time-dependent image of the plasma emissivity, $\eps(\xbf,t)$.
The emisivity is advected by an unknown, time-dependent velocity,
$\ub(\xbf,t)$, so we postulate a stochastic model:
$$
\part_t \eps + \ub\cdot \nabla \eps =  \kappa_o \Delta \eps +\xi_1 \ .
\eqno (6.5)$$
To apply the extended Kalman smoother to estimate both  
 $\eps(\xbf,t)$ and $\ub(\xbf,t)$, we augment the system with
$$\part_t \ub  = \nu\Delta \ub +\xi_2 \ , \eqno (6.6)$$ 
$$\nabla \cdot \ub= 0 \ .$$
To simplify the model, we have dropped the forces from Eq.~(6.6).
Both $\kappa_o$ and $\mu$ can be given by neoclassical theory or 
can be estimated. The advantages of this approach are: a) the velocity
field is estimated from the time-dependent emissivity; 
b) by using many time slices of
the tomography data simultaneously, coupled by Eqs. (6.5-6),  we reduce the
error in the estimate of $\eps(\xbf,t)$. 

\ns \ni 
{\bf VII. Discussion}

\medskip

In this article, we applied the augmented Kalman smoother to inverse problems 
in partial differential equations. The stochastic forcing term
in the evolution equations represents model error. 
This stochastic forcing reduces the chances of filter divergence.
Our approach is common for finite dimensional engineering problems,
but we are unaware of any previous work which uses the augmented Kalman 
smoother to estimate unknown coefficients in 
distributed systems of partial differential equations.

The nonlinear coupling term, $\kaptlh \nabla \Ttlh$ introduces non-Gaussianity
into the system. In using the extended iterated Kalman smoother, we neglect
the non-Gaussian part of the estimation problem. A rigorous alternative
is consider the full nonlinear filtering problem using the Zakai
equation in function space. Unfortunately, this function space estimation 
approach is  computationally infeasible.

Thus we use the extended  Kalman smoother with two modifiications to
reduce the non-Gaussianity and numerical ill-conditioning.
First, we add a regularizing term, $\mu_1 \part_x^4 T(\xbf,t)$,
to damp out higher-order oscillations 
which are below the resolution threshold, and thereby aid in convergence. 
Second, we update the Kalman smoother by adding
a temporally smoothed version of $\Ttlh$. The kernel smoother reduces 
the variance of $\Tbr$ and corresponds to the probabilistic
expectation. 
We update $\Tbr$ by averaging the heat flux, $ \kaph\nabla \Th$, 
and then solving Eq.~(5.3) for $\Tbr$.

\medskip

{\it Acknowledgment}

The author has benefited from discussions with G. Papanicolaou and B. Kohn.
The referee's comments are gratefully acknowledged.
This work was performed under U.S. Department of Energy Grant No.
DE-FG02-86ER-53223.

\begin{center}
{\bf REFERENCES}
\end{center}

\begin{enumerate}

\item S.I.~Aihara,   
{\it S.I.A.M.~J.~of Control and Opt.}, {\bf 30}, 745 (1992). 

\item B.D.O.~Anderson and J.B.~Moore,  {\it Optimal Filtering.}
(Prentice-Hall, New Jersey 1979).
 

\item  A.V.~Balakrishnan,  {Parameter estimation in stochastic 
differential systems: theory and applications.} In {\it Developments
in Statistics.} P.R.~Krishnaih, ed., 
(Academic Press, New York 1978).



\item H.T.~Banks  and P.D.~Lamm,   
{\it  IEEE Trans.~on Automatic Control} {\bf 30}, 386 (1985).  


\item J.V.~Beck, B.~Blackwell,   and  C.R.~St.~Clair, 
{\it Inverse Heat Conduction - Ill-posed problems.}
(Wiley-Interscience Publications, New York 1985).~

\item A.~Bensoussan, 
 {\it Filtrage Optimal des Syst\`{e}mes Lin\'{e}aires.}
(Dunod, Paris 1971).


\item J.~Blum,  
{\it Numerical Simulation and Optimal Control in Plasma Physics.}
(Wiley/Gauthiers-villars, Chichester-New York 1989).


\item A.E.~Bryson Jr.~and Y.C.~Ho,  {\it Applied Optimal Control}.
(Blaisdel Publishing Co., New York 1969).

\item S.E.~Cohn  and D.F.~Parrish,  
{\it Monthly Weather Review}, {\bf 120}, 1757 (1991).

\item R.~Curtain,  
{\it S.I.A.M.~Review}, {\bf 17}, 395  (1975). 

\item A.~Fiacco and G.~McCormick,  {\it Nonlinear Programming: Sequentially
Unconstrained Minimization Techniques}, (J.~Wiley, New York 1968).


\item M.~Ghil,   
{\it Dynamics of Oceans and Atmospheres}, {\bf 13}, 171 (1989).

\item R.~Glowinski,
{\it J.~of Comp.~Physics}, {\bf 103}, 189 (1992).      

\item R.J.~Goldston,  D.C.~McCune,  H.H.~Towner,  et al., 
{\it J.~of Comp.~Physics}, {\bf 43}, 61 (1981).

\item G.C.~Goodwin  and R.L.~Payne, {\it Dynamic System Identification:
Experimental Design and Data Analysis}. 
(Academic Press, New York 1977).

\item R.S.~Granetz  and  P.~Smeulders, 
{\it Nuclear Fusion}, {\bf 28}, 457 (1988). 


\item  K.-H.~Hoffman and J.~Spreckels,   
{\it Num.~Funct.~Anal.~and Optim.}, {\bf 7}, 157 (1984). 

\item S.C.~Jardin,  N.~Pomprey,  and J.~Delucia, 
{\it J.~of Comp.~Physics}, {\bf 66}, 481 (1986).      

\item A.H.~Jazwinski,   {\it Stochastic Processes and
Filtering Theory}.~(Academic Press, New York 1970).

\item  C.E.~Kessel  and M.A.~Firestone,  
{\it  IEEE Trans.~on Plasma Science}, {\bf 19}, 29 (1991). 


\item P.E.~Kloeden and E.~Platen, 
{\it Numerical Solution of Stochastic Differential Equations},
(Springer-Verlag, Berlin 1992).

\item C.~Kravaris and J.H.~Seinfeld,  
{\it S.I.A.M.~J.~of Control and Opt.}, {\bf 23}, 217 (1985). 

\item  M.~Kress and K.S.~Riedel,  
{\it Journal of Comp.~Physics}, {\bf 83}, 237  (1989).

\item  P.D.~Lax, and R.D.~Richtmyer,  
{\it  Comm.~Pure and Appl.~Math.}, {\bf 9},  267 (1956). 

 


\item S.~Omatu and J.H.~Seinfeld,  
{\it Distributed Parameter Systems, Theory and Applications.}
(Oxford Science Publications, Oxford 1989). 


\item K.S.~Riedel, E.~ Eberhagen, O.~Gruber,  et al., 
{\it Nuclear Fusion}, {\bf 28}, 1509  (1988).

\item K.S.~Riedel,   
{\it Automatica}, {\bf 29}, 779  (1993). 




\item P.K.C.~Wang,   Identification problems in plasma physics.
{\it Lecture Notes in Control and Information Sciences} {\bf 1}, 424-445,
(Springer-Verlag, Berlin 1978).

\item S.J.~Wright, Interior  point methoods for optimal control of
discrete time systems. Tech. Rep.~MCS-P229-0491, Argonne Nat.~Lab.,
Argonne, IL (1991).


\end{enumerate}{}

\ns \noindent
{\bf Appendix A: Variational Formulations}

\medskip

For completeness, we state  two variational formultions of the
Kalman smoother. The  variational formulations are useful from a
Bayesian perspective and may yield higher-order discretizations. 
For certain variational formulations [6,22], we need to assume that
$\Fb(\thbr(t),t)$ maps  the Hilbert space, $V \subset L^2(R^n), $
into its  dual space, $V'$, such that (i) the map from
$V$ to $V'$ is bounded in the operator norm,
(ii) the map is coercive:
$(\Fb(\thbr(t),t)\ubf(t),\ubf(t))_{L^2} + c ||\ubf||_{L^2}^2 \ge
\alpha ||\ubf||_V^2$. We  assume that these properties are satisfied 
not only at $\th =\thbr$,
but for all $\th$ in a compact, convex region, $\Omega$, with 
$\thbr \in \Omega$.
For parabolic problems, $V$ is the space of functions with square integrable
derivatives and the appropriate boundary conditions and 
$||u||_V^2 \equiv \int  |\nabla u|^2 +u^2  dx^n$.
For the inverse heat conductivity problem of Eq.~(1.4), the  hypotheses
are satisfied provided that the diffusivity remains  bounded from above 
and below.

The constrained least squares formulation of the Kalman smoother is 
$$\ell(\ub,\lambda,\xi|\yb,\Fb,\Qb) \equiv
 \int_{0}^{t_f} dt 
[\yb_{t} - \Hb_t \ub_t ]^{*} \Rb^{-1} [\yb_{t} - \Hb_t \ub_t ] \ + $$
$$\int_{0}^{t_f} dt \int d\xbf \xi_t^{*} \Qb_{t}^{-1}\xi_t
+ \int_{0}^{t_f} dt \int d\xbf 
\lambda_t [\dot{\ub}_{t} - \Fb_t \ub_t - \sb_t - \xi_t]
,\eqno (A1)$$
which yields to the backward time adjoint equation for the Lagrange
multiplier:
$$
\partial_t{\lambda} (t) = -\Fb_t^*  \lambda -
\Hb^{*} \Rb^{-1} [ {\yb}_t - \Hb_t \hat{\ub} (t|t_f) ] \ .  \eqno (A2)
$$
When $\Qb_{\GB,t}$ is also positive definite and trace class, the
least squares functional is
$\ell(\ub|\yb,\Fb,\Qb) \equiv$ 
$$\int_{0}^{t_f} dt \left[  
(\yb_{t} - \Hb_t \ub_t )^{*} \Rb^{-1} (\yb_{t} - \Hb_t \ub_t ) +
\int d\xbf (\dot{\ub}_{t} - \Fb_t \ub_t - \sb_t)^{*}
\Qb_{\GB,t}^{-1}(\dot{\ub}_{t} - \Fb_t \ub_t - \sb_t) \right]
.\eqno (A3)$$
Equation (A3) is the continuous time analogy of Eq.~(B10).

\ns \ni
{\bf Appendix B. Discrete Time Kalman Filter-Smoothers} 

\medskip

We review discrete Kalman filters and smoothers;
more extensive presentations are given in the textbooks by
Jazwinski [19] and Anderson \& Moore [2]. 

\medskip
\ni
A) Discrete Kalman Filter-Smoothers

We consider the discrete linear state space model:
$$
\ub_{i+1} = \Phi (i+1,i) \ub_i + \GB_i{\wb}_{i} +\sb_i
,\eqno (B1)$$
$$
\yb_{i} = \Hb_i \ub_i + {\eps}_i
,\eqno (B2)$$
where $\ub_i$ is the state vector of dimension $N$,
$\yb_i$ is the measurement vector of dimension $m$, and
$\sb_i$ is the known source vector of dimension $N$.
$\Phi (j,i)$ is the $N \times N$ nonsingular deterministic part of
the map from time $i$ to time $j$. The system noise, $\wb_i$,
is assumed to be an $r$-dimensional white Gaussian processes with
covariance $\Qb_i$.
The measurement noise, ${\eps}_i$, is a $m$-dimensional white Gaussian sequence
with nondegenerate covariance $\Rb_i$. The $m \times N$ measurement-evaluation
matrix, $\Hb_i$, maps the state vector, $\ub_i$, onto
the deterministic part of the measurements. To simplify
the notation, we define the $N\times N$ matrices: $\Fb_i \equiv \Phi (i+1,i)$, 
$\Qb_{\GB,i} \equiv \GB_i \Qb_i \GB_i^{*}$ and  
$\Jb_{i} \equiv \Hb_i^{*} \Rb_i^{-1} \Hb_i$.


The standard Kalman filter estimates the state vector, 
$\hat{\ub}(i|j)$, at time $i$ given
the measurements, $\yb_1 , \ldots , \yb_j$ up to time $j$
by the time evolution update:
$$
\hat{\ub}(i+1|i) = \Fb_i \hat{\ub}(i|i)  +\sb_i \ .\eqno (B3)
$$
The covariance, $\Pb(i|j)$,  of the estimate, $\hat{\ub}(i|j)$,
evolves as
$$\Pb(i+1|i) = \Fb_i \Pb(i|i) \Fb_i^{*}  + 
\Qb_{\GB,i}  \ . \eqno (B4)$$
We assume that $\hat{\ub}(0|0)$ and $\Pb(0|0)$ are given.
The measurement update is
$$
\hat{\ub}(i|i) = \hat{\ub}(i|i-1) + \Kb_i ( \yb_i -
\Hb_i \hat{\ub}(i|i-1) )
,\eqno (B5)$$
$$
\Pb(i|i)^{-1} = \Pb(i|i-1)^{-1} + \Hb_i^{*} \Rb_i^{-1} \Hb_i 
,\eqno (B6)$$
where $\Kb_i$ is the $N \times m$ Kalman gain:
$$
\Kb_i = [\Pb(i|i-1)^{-1} + \Hb_i^{*} \Rb_i^{-1} \Hb_i ]^{-1} \Hb_i^{*} \Rb_i^{-1}
= \Pb(i|i) \Hb_i^{*} \Rb_i^{-1}
. \eqno (B7)$$
Following the Rauch-Tung-Striebel (R.T.S.) formulation (Ch.~13.2 of Ref.~8), 
we divide the Kalman smoother into 
a forward Kalman filter followed by a backward smoother correction: 
$$
\hat{\ub}(i|N_f) = \hat{\ub}(i|i)+
\Pb(i|i) \Fb_i^{*} \Pb^{-1}(i+1|i) 
\left( \ubh(i+1|N_f)- \ubh(i+1|i) \right)
, \eqno (B8) $$
$ \Pb(i|N_f) = \Pb(i|i) + $
$$
\Pb(i|i) \Fb_i^{*} \Pb^{-1}(i+1|i) 
\left[ \Pb(i+1|N_f)- \Pb(i+1|i) \right]  
\Pb^{-1}(i+1|i) \Fb_i\Pb(i|i) 
.\eqno (B9) $$

\vspace{.21in}
\noindent
B) Least squares formulation  of the Kalman smoother

The fixed interval Kalman smoother is the least squares and maximum
likelihood estimator of $\ub_i$, 
given the measurements, $\yb_1 \ldots \yb_{N_f}$,
where the final measurement time is $N_f$. 
The Kalman smoother is a statistical estimation problem of very large
dimension, $N  N_f$.

The generalized least squares functional, $\ell(\ub|\yb,\Fb,\Qb)$, 
for the Kalman smoother is $\ell(\ub|\yb,\Fb,\Qb) \equiv$ 
$$ \sum_{i=1}^{N_f}
(\yb_{i} - \Hb_i \ub_i )^{*} \Rb^{-1} (\yb_{i} - \Hb_i \ub_i ) +
\sum_{i=1}^{N_f} (\ub_{i+1} - \Fb_i \ub_i - \sb_i)^{*}
\Qb_{\GB,i}^{-1}(\ub_{i+1} - \Fb_i \ub_i - \sb_i)
,\eqno (B10)$$
where $\Qb_{\GB,i}^{-1} \equiv (\GB_i \Qb_i \GB_i^{*})^{-1}$ 
The least squares functional has a Bayesian interpretation:
$ \sum_{i=1}^{N_f} (\ub_{i+1} - \Fb_i \ub_i - \sb_i)^{*}
\Qb_{\GB,i}^{-1}(\ub_{i+1} - \Fb_i \ub_i - \sb_i)$
is the argument of the {\it a priori} probability density of the
stochastic system, and $\sum_{i=1}^{N_f}
(\yb_{i} - \Hb_i \ub_i )^{*} \Rb^{-1} (\yb_{i} - \Hb_i \ub_i )$
represents the conditional probability density.
Thus $\ell(\ub|\yb)$ is the argument of the {\it a posteori} 
probability density. 
In the Bayesian interpretation, $\ubh_i$ is the weighted sum of the
{\it a priori} estimate and a new independent estimate from the measurements,
$\yb_1 \ldots \yb_{N_f}$.

In dynamic programming, $\ub$ is estimated by minimizing  
$\ell(\ub|\yb,\Phi,\Qb)$ directly.
Differentiating $\ell(\ub|\yb,\Fb,\Qb)$
with respect to $\ub_i$ yields the estimation equations for 
$\hat{\ub}(i|N_f)$:
$$
-\Qb_{\GB,i-1}^{-1} \Fb_{i-1} \hat{\ub}(i-1|N_f) \  +
\left( \Fb_i^{*} \Qb_{\GB,i}^{-1} \Fb_i \ +
\Qb_{\GB,i-1}^{-1} +  \Hb_i^{*} \Rb_i^{-1} \Hb_i \right)\hat{\ub}(i|N_f)
$$ $$
- \Fb_i^{*} \Qb_{\GB,i-1}^{-1}  \hat{\ub}(i+1|N_f)
= -\Fb_i^{*} \Qb_{\GB,i}^{-1} \sb_i + \Qb_{\GB,i}^{-1} \sb_{i-1} 
+ \Hb_i^{*}\Rb_i^{-1}\yb_i 
\ . \eqno (B11a)$$
We decompose $\ub_i$ into a deterministic component, $\ubb_i$,
and the stochastic component, $\ubt$. Equation (B9a) reduces to
$\ubb_{i+1} - \Fb_i \ubb_i - \sb_i$, and 
$$
-\Qb_{\GB,i-1}^{-1} \Fb_{i-1} \tilde{\ub}(i-1|N_f) \  +
\left( \Fb_i^{*} \Qb_{\GB,i}^{-1} \Fb_i \ +
\Qb_{\GB,i-1}^{-1} +  \Hb_i^{*} \Rb_i^{-1} \Hb_i \right)\tilde{\ub}(i|N_f)
$$ $$
- \Fb_i^{*} \Qb_{\GB,i-1}^{-1}  \tilde{\ub}(i+1|N_f)
= \Hb_i^{*}\Rb_i^{-1}(\yb_i -\Hb_i \ubb)
\ . \eqno (B11b)$$
 Equation (B11b) is a coupled set of $NN_f$ equations with a symmetric, block
tridiagonal structure. The inverse of the 
{\it a posteori} covariance matrix, $\Sib$, is defined by
$$\Sib_{i,i}^{-1} \equiv
\left( \Fb_i^{*} \Qb_{\GB,i}^{-1} \Fb_i \ +
\Qb_{\GB,i-1}^{-1} +  \Hb_i^{*} \Rb_i^{-1} \Hb_i \right) \ ,
$$
$$ \Sib_{i,i-1}^{-1} \equiv
- \Qb_{\GB,i-1}^{-1} \Fb_{i-1} \  ,
\Sib_{i,i+1}^{-1} \equiv - \Fb_i^{*} \Qb_{\GB,i}^{-1}
\ . \eqno (B12)$$
The block tribanded structure enables the Kalman smoother equations
to be solved using
a forward sweep and then a backward sweep of the equations. 
{\it The forward-backward sweeps of the R.T.S. smoother correspond 
to the standard algorithm for solving block tribanded matrices.} 


If the estimated
covariance, $\Pbf(t)$, is small relative to $\Rbf$ (more precisely,
if $\Pb(i|i-1)^{-1} >> \Hb_i^{*} \Rb_i^{-1} \Hb_i$ (in the discrete  case)), 
the continuous measurement time filter may be used.  
If the measurement times are slower than the characteristic 
evolution time, the time map, $\Phi(i+1,i)$, between
the $i$th and $i+1$th measurements is evaluated by the composition
of many time steps.

\ns \ni
{\bf Appendix C. Combining the Kalman Smoother with 
Smoothness Penalty Functions} 

\medskip

A number of researchers have examined parameter estimation  when
$\kappa(\xbf)$ is an unknown, deterministic  function of space
($\Qb_{T}\equiv 0$ and $\Qb_{\th}\equiv 0$). 
To regularize
the infinite dimensional minimization, either $\kappa(\xbf)$ is required
to be in a compact convex subset of smooth functions where 
$0 < \kappa_L \le \kappa(\xbf) \le \kappa_U$ [3,4], or 
a smoothness penalty function is added . Thus the penalized least squares
problem is to minimize
$$\ell(\ub,\kapbr|\yb) \equiv
 \int_{0}^{t_f} dt 
(\yb_{t} - \Hb \ub_t )^{*} \Rb^{-1} (\yb_{t} - \Hb \ub_t ) + 
\delta ||\thbr||_{}^2
,\eqno (C1)$$
over initial conditions and over $\thbr = ln(\kappa(\xbf))$ in a convex set with
$0 < \kappa_L \le \kappa(\xbf) \le \kappa_U$ 
and $||\theta||_{s}^2$ is a smoothness penalty. 
Equation (C1) can be interpreted as the Bayesian $a$ $posteori$ likelihood
given a smoothness prior. Alternatively, Eq.~(C1) can be interpreted as
an approximate likelihood where the penalty term induces a bias error in
order to reduce the variance of the estimate by damping the high frequency 
modes.
  
The convergence of discretized approximations is proven in [4,6,22].
Kravaris \& Seinfeld [22] give general conditions when penalized
least squares functionals have an unique minimum. Recently, Aihara [1]
has applied this analysis to the case of stochastic forcing in the temperature 
equation with $\kappa(\xbf)$ an unknown, deterministic  function of space. 

Banks and Lamm [4] 
consider the case of deterministic dynamics with time-dependent diffusivities.
They show that
the temporally and spatially discretized parameter estimates converge to
the correct parameter estimates of the continuous problem as the discretization
becomes finer.
Alternately, the results of  Aihara can be generalized to time-dependent
diffusivities. To ensure the existence of a unique minimizer of the
loss functional, $\kappa(\xbf,t)$ needs to be restricted to a compact set
or a smoothness penalty in both time and space, such as  
$$||\thbr||_{s}^2 =
\int_{0}^{t_f} dt \int d\xbf 
\left[ |\partial_t^2 \thbr(\xbf,t)|^2 + |\Delta \thbr(\xbf,t)|^2 \right]
,\eqno (C2)$$
must be applied. Similarly, our augmented Kalman smoother formulation adds 
a penalty term of the form 
$$\int d\xbf (\dot{\thtl}_{t} - \Mbf \thtl)^{*}
\Qb_{\th}^{-1}(\dot{\thtl}_{t} - \Mbf \thtl)
\ .\eqno (C3)$$
 As $\Qb_{\th}$ increases, 
the importance of the most recent measurements in determining 
the instantaneous value of $\thbr$ increases. 
Thus the  stochastic forcing, $\xi_{\thtl}$, reduces 
the tendency of the smoother to diverge.

Both Eq.~(C2) and Eq.~(C3) penalize
against rapid, unphysical variations in the parameter estimates by adding 
an $a$ $priori$ smoothness density to the likelihood function.
Our augmented Kalman filter approach has several advantages.
For given $\Tbr$ and $\kapbr$, the Kalman smoother is the exact 
minimizer of the likelihood function with respect to $\Ttl$ and $\thtl$.
In contrast, the standard formulation of the penalized likelihood approach
of Eq.~(C1-C2) requires many iterations of a steepest descent algorithm
to minimize with respect to $\kappa$ [1,22]. 
Furthermore, the computation of the
gradient of Eq.~(C1) with respect to $\kappa$ is extremely complicated [1].
The computational simplicity of our formulation arises because $\thtl$
and $\Ttl$ enter quadratically in our likelihood functional.

The two approaches have  slightly different functions.
The smoothness penalty damps all rapid variation while the prior
of Eq.~(C3) damps large innovations, $\wbf$. 
The two approaches may be combined by adding a prior distribution
of $ ||\Ttl||_{}^2 + ||\thtl||_{}^2$ to the standard variational
formulation  of Eq.~(A1). 
As an alternative to the smoothness penalty, we prefer to add an
artificial hyperdiffusivty, $\mu \Delta \Delta \Ttl$, to the
evolution equations. The hyperdiffusivity damps the high frequency
modes as does the smoothness penalty. The addition of a hyperdiffusivity
is easy to implement numerically and lets us  remains in the Kalman
smoother framework. The artificial hyperdiffusivity induces
a systematic bias error in exchange for reducing the ill-conditioning.
Equation (4.2) quantifies this bias versus variance tradeoff.


\newpage
  \noindent
{\bf Appendix D.  Nearly Block Diagonal Implementation of the Kalman Smoother}

\medskip

In [27], 
we showed that the computational cost and numerical
ill-conditioning of the Kalman filter/smoother can be noticeably reduced
when the stochastic system, including $\Hb^{*}\Rb^{-1}\Hb$ and $\Qbf$, 
is nearly block diagonal (N.B.D.), and
first-order approximations are sufficient. In our case, the computational
effort decreases by a factor of $m$. 
To apply the N.B.D. approximation to Eqs. (3.6-8),
we assume that the conductivity is nearly constant, and expand
$\th(r,t) =\th_0 +\th_1(r,t)$. 

Spatial aliasing strongly couples the Fourier modes with 
$k-k' = 0$ mod $m$. To preserve the block diagonal structure for
$\Hb^{*}\Rb^{-1}\Hb$, we reorder the indices in the state vector, $\ub$.
The $k$th block of the reordered $\ub$ is 
$T_k,\th_k,T_{k+m},\th_{k+m} \ldots$
We expand $\Hb^{*}\Rb^{-1}\Hb$ about a block diagonal matrix,
and include the corrections due to unequal variances 
($\sigma^2_k \ne \sigma^2_{k'}$), and nonuniformly distributed
measurement locations ($x_{\ell}\ne \pi{(2\ell-m-1)\over m}$).
The reordered and expanded system is then solved using the first-order
approximation to the iterated Kalman smoother.

 \ns \noindent
{\bf Appendix E.  Models for Stochastic Parameter Evolution}

\medskip

Ideally, the model  error  covariances, $\Qb_{\GB}$ and $\Qb_{\th}$,
are given $a$ $priori$ or are estimated from the residuals.
$\Pbf(t,s)$ is the covariance of the estimates, $\ubh$ {\it conditional
on the measurements}. 
We examine $\Cbf(t,s)$, the covariance of 
$\ubf^{*} = (\Ttl,\thtl)^{*}$ in the {\it absence of measurements},
In practice, we often have a better understanding of the size and 
autocorrelation  time of $\Cbf_{}$ than of the stochastic model parameters,
$\Qbf$ and $\Mbf$.

We now determine $\Cbf_{}(s,t)$ in terms of $\Fbf$ and $\Qbf$
when  $\Fbf$ and $\Qbf$ are time-independent. The covariance evolves
$\partial_t \Cbf(s,t)  = \Fbf \Cbf + \Cbf\Fbf^{*} + \Qbf$.
We assume that all
of the eigenvalues of $\Fbf$ have negative real parts, and therefore
$\ubf$ has a stationary covariance: $\Cbf_{}(s,t) = \Cbr(s-t)$, where
$$\Cbf_{}(t,t) = \Cbr_{}(0) \equiv 
\int_0^{\infty} exp{(t'\Fbf)} \Qbf_{} exp{(t'\Fbf^{*})} dt'
\ ,\eqno (E1)$$
$$\Cbf(s,t) = \Cbr_{}(s-t) = \Cbr_{}(0) exp{((t-s)\Fbf^{*})}
\ \ \ {\rm for} \ s  \le t .\eqno (E2)$$
If we use a basis where $\Fbf$ is diagonal, Eqs. (E1-2) reduce to
$$C_{i,j}(s,t) =  
{e^{(t-s)\lambda_j}Q_{i,j} \over (\lambda_i + \lambda_j)}
\ \ \ {\rm for} \ s \ \le t 
\ .\eqno (E3)$$
Thus if $\Qbf$ and $\Fbf_{}$ are simultaneously diagonalizable,
then the different eigenvector directions are uncoupled and evolve
independently. We  will generally choose $\Qbf$ to be diagonal in the
basis of $\Fbf$ eigenfunctions.

When the autocorrelation time of the fluctuating $\Ttl$ and $\thtl$
fields is short in comparison with the characteristic time scale
for $\Tbr$ and $\kapbr$ evolution, we can treat the fluctuations as 
quasistationary. Since Eq.~(1.4) has a block upper triangular structure,
the eigenfunctions of the frozen time  version of Eq.~(1.4) are of two
types. The first class are eigenfunctions of the form:
$(T^1_j(\xbf),\th^1_j(\xbf))^{*} = (T^1_j(\xbf),0)^{*}$ where
$T^1_j(\xbf)$ is an eigenfunction of 
$\nabla\cdot \kapbr(\xbf,t) \nabla T^1_j(\xbf) = \lambda^1_j T^1_j(\xbf)$.
The second class is based on the eigenfunctions of Eq.~(1.4b),
$(T^2_j(\xbf),\th^2_j(\xbf))^{*}$ where
$\mu_{2}\Delta \th^2_j(\xbf) = \lambda^2_j \th^2_j(\xbf)$.
Thus the eigenfunctions of the perturbed temperature operator,
$\nabla\cdot \kapbr(\xbf,t) \nabla$, do not couple to $\thtl$,
but the eigenfunctions of $\mu_{2}\Delta$ couple to $\Ttl$.
Correspondingly, we divide the covariance, $\Cbf$ into
$$\Cbf = \left( 
\begin{array}{c} \Cbf^1_{T,T} \\ 0 \end{array}
\begin{array}{c} 0 \\ 0 \end{array} 
\right) \  + \ \left(
\begin{array}{c} \Cbf^2_{T,T} \\ 0 \end{array}
\begin{array}{c} \Cbf^2_{T,\th} \\ \Cbf^2_{\th,\th}  \end{array}
\right) .\eqno (E4)$$
As the mode number, $k$ increases, the cross-term, $\Cbf_{T,\th}$,
decreases relative to the  diagonal terms as $1/k$.
The asymptotic decoupling of the $\Ttl$ equation and the 
$\thtl$ equation implies that the variation of $\thtl$ becomes
increasingly difficult to identify as the wavelength of $\thtl$
variation decreases.


\end{document}